\documentclass[11pt]{article}
\usepackage[margin=1in]{geometry}
\usepackage{graphicx}
\usepackage{amsmath}
\usepackage{booktabs}
\usepackage{caption}
\usepackage{subcaption}
\usepackage{hyperref}
\usepackage{xcolor}
\usepackage{natbib}
\usepackage{float}

\hypersetup{colorlinks=true, urlcolor=blue!60!black, linkcolor=black, citecolor=black}

\graphicspath{{figures/}}

\title{\vspace{-2em}\textbf{AlphaEarth Satellite Embeddings for Modelling Climate Sensitive Diseases Towards Global Health Resilience}}

\author{\small Usman Nazir, I-Han Cheng, Sara Khalid \\
\small Planetary Health Informatics (PHI) Lab, University of Oxford \\
\small {\{usman.nazir, sara.khalid\}@ndorms.ox.ac.uk} \\
i-han.cheng.24@ucl.ac.uk\\
}
\date{}
\begin{document}
\maketitle
\thispagestyle{empty}
\vspace{-1.5em}

\begin{abstract}
\noindent 
Malaria, childhood acute respiratory infection, and child undernutrition together account for over two million deaths annually in children under five, with the burden concentrated in low and middle-income countries where climate variability directly modulates transmission, exposure, and nutritional outcomes (\cite{gbd2019, who2023malaria, unicef2023}) . Despite this, routine health surveillance in these settings remains sparse and reactive. Satellite-derived representations of the Earth's surface offer a scalable, low-cost complement to traditional covariates, yet their utility as predictors of population health outcomes is poorly characterised \citep{bhatt2015}. We summarise findings from three independent studies that evaluate AlphaEarth Foundations 64-dimensional satellite embeddings as predictors of population health outcomes, with a focus on vulnerable population groups e.g. children and populations in low-resource settings. The studies span two health domains: infectious disease (malaria, respiratory infection) and stunting.\footnote{Following common usage in the child-nutrition community, we use ``stunting'' as the umbrella term for chronic child undernutrition. The specific outcome modelled in Case 3 is the Weight-for-Height Z-score (WHZ), which is the continuous variable from which the binary \emph{wasting} indicator (WHZ $<$ $-2$) is derived. Stunting (low height-for-age) and wasting (low weight-for-height) are related but distinct dimensions of undernutrition; our pipeline is agnostic to which of the two is selected as the target.} In each study, embeddings provide meaningful predictive value when merged at sufficient spatial granularity: (i) in malaria case prediction across Nigeria, AlphaEarth Embeddings yields consistent per-region $R^2$ gains; (ii) in childhood acute respiratory infection (ARI) prediction across 11 DHS countries, embeddings increase pooled $R^2$ from 0.157 to 0.206, an effect that is robust across three tree-based estimators; (iii) in stunting prediction across 35 countries, a country-level prototype is---as expected---neutral, because country-broadcast embeddings are collinear with country fixed effects. The stunting case is currently blocked on DHS cluster-level coordinate access; we view this as the next and most informative experiment. We close with a request: direct access to the Google Earth AI foundation-model suite, including Population Dynamics embeddings that incorporate health indicators, would substantially accelerate this line of work across all three outcomes.
\end{abstract}

\begin{figure}[H]
\centering
\begin{subfigure}[t]{0.45\textwidth}
    \centering
    \includegraphics[width=\textwidth]{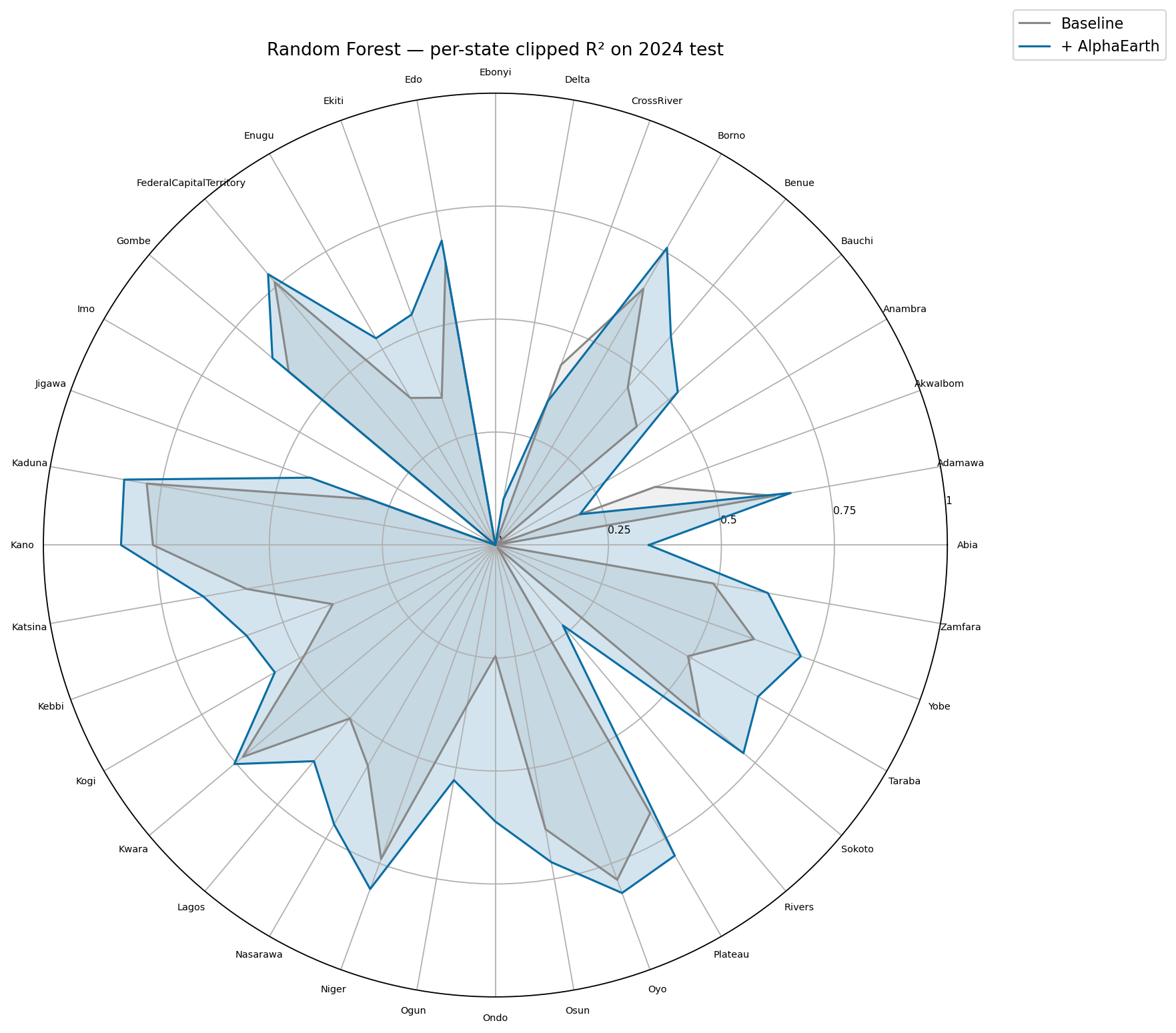}
    \caption{\textbf{Per-region 2024 test $R^2$ across Nigerian states.} Each spoke is one state; the inner polygon is the climate-only baseline and the outer polygon is the same model with the 64-dim AlphaEarth fingerprint appended. The outer polygon strictly dominates on every spoke, indicating that the lift is geographically uniform rather than driven by a few high-burden states.}
    \label{fig:case1_radar_rf}
\end{subfigure}
\hfill
\begin{subfigure}[t]{0.45\textwidth}
    \centering
    \includegraphics[width=\textwidth]{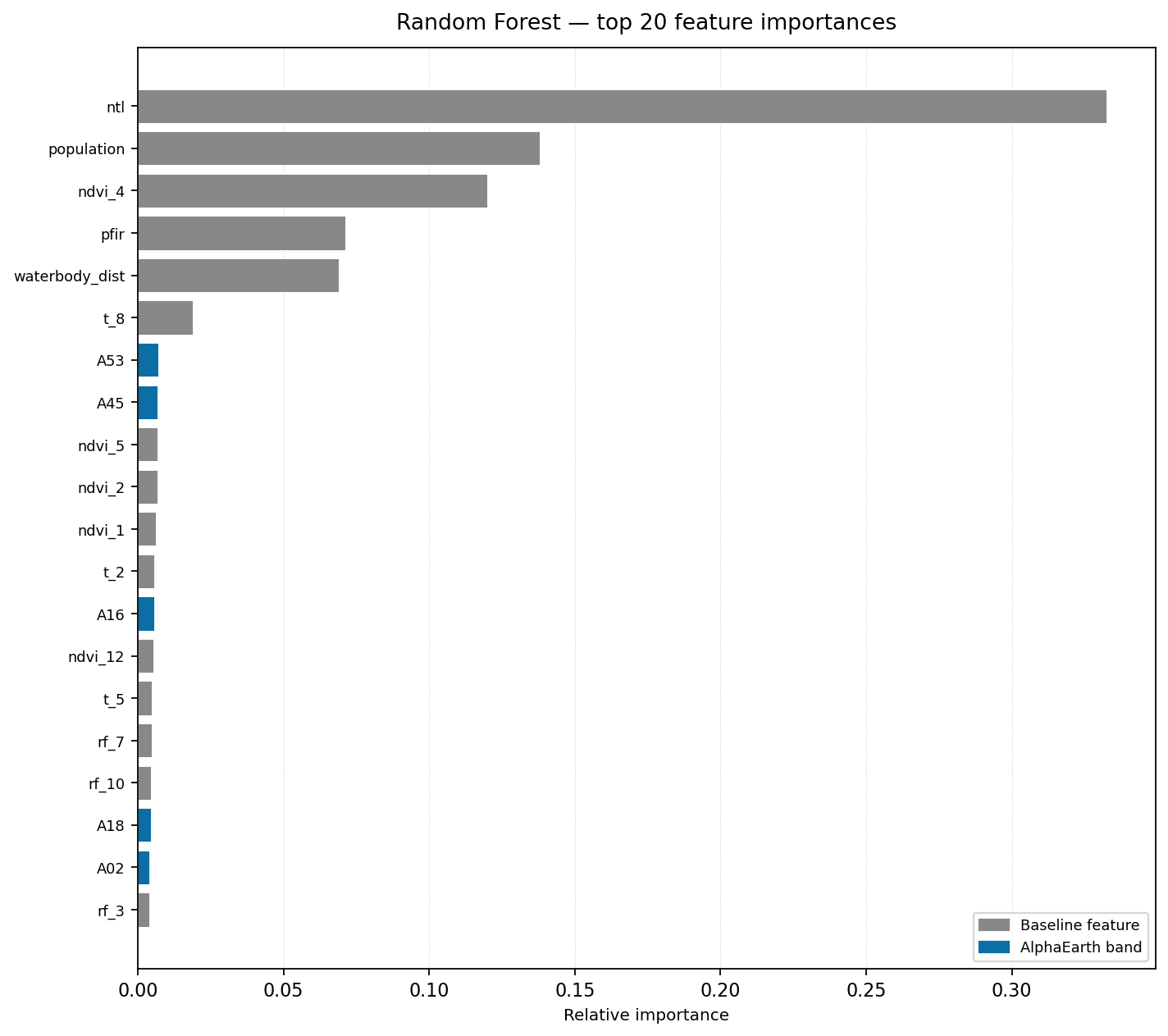}
    \caption{\textbf{Top-20 Random Forest feature importances.} Bars are coloured by feature group (baseline climate / contextual vs. AlphaEarth band $A_i$).}
    \label{fig:case1_bar}
\end{subfigure}
\caption{\textbf{Case 1 --- Malaria case prediction in Nigeria (NMEP, 2000--2024; train 2000--2023, test 2024).} AlphaEarth embeddings provide a geographically uniform $R^2$ gain (left) and emerge as the dominant feature group in the importance decomposition (right), supporting the interpretation that static landscape structure carries malaria-transmission signal not captured by monthly climate covariates.}
\label{fig:case1}
\end{figure}

\section{Motivation}

Malaria, acute respiratory infection, and undernutrition are among the leading causes of child mortality globally, collectively responsible for a disproportionate share of the disease burden in sub-Saharan Africa and South Asia ~\citep{gbd2019}. Climate variability is a well-established driver of all three: temperature and rainfall govern mosquito breeding and Plasmodium falciparum transmission~\citep{mordecai2019, who2023malaria}, ambient air quality and indoor pollution modulate respiratory infection risk, and seasonal drought and flood cycles affect food security and nutritional status~\citep{unicef2023}. As climate change intensifies these exposures, the need for scalable, near-real-time environmental monitoring tools for health is increasingly urgent \citep{watts2022}. 

Child and population health in LMICs is driven by environmental and structural factors, such as climate, vegetation, infrastructure, land cover, water  which  are difficult to capture with traditional survey covariates. Geolocated Demographic and Health Survey (DHS) data provides cluster-level outcomes at 2--5 km spatial precision, creating a natural opportunity to test whether learned representations of the Earth's surface carry predictive signal for health outcomes.

We evaluate AlphaEarth Foundations V1 (64-dim, 10-m annual unit-norm embeddings via Google Earth Engine) on three distinct LMIC outcomes spanning two health domains:
\begin{itemize}
\item \textbf{Malaria} (vector-borne infectious disease): Nigeria, 2000--2024.
\item \textbf{Childhood ARI} (respiratory infectious disease): 11 DHS countries, 2017--2022.
\item \textbf{Child undernutrition (WHZ)} (chronic / stunting): 35 DHS countries, 2015+.
\end{itemize}

Each study uses an appropriate baseline (climate-only, pollution-only, or regression-based tabular respectively), and tests whether adding AlphaEarth embeddings improves test-set prediction. The three case studies are presented sequentially below.

\section{Case 1 --- Malaria Prediction, Nigeria}

\paragraph{Setup.} Annual malaria case counts from Nigeria's National Malaria Elimination Programme (NMEP) surveillance across 28{,}209 locations, 2000--2024, linked to monthly ERA5-Land climate (temperature, rainfall, NDVI), contextual covariates, and a static AlphaEarth fingerprint per location. Training is performed on 2000--2023 and held-out evaluation on 2024, giving a one-year forward gap that stresses the model's ability to generalise beyond the training window. \emph{Plasmodium falciparum} prevalence (\texttt{pfir}) is explicitly excluded from inputs to avoid circular prediction.

\paragraph{Result.} On the 2024 test year, adding AlphaEarth raises test $R^2$ from 0.201 to 0.245. All Nigerian regions show positive $\Delta R^2$ in Figure~\ref{fig:case1_radar_rf}  (Also see Figure~\ref{fig:case1_radar_resnet} in appendix~\ref{case1_resnet}).

\section{Case 2 --- Childhood ARI, 11 DHS Countries}
\paragraph{Setup.} 9{,}271 DHS clusters across India, Nigeria, Pakistan, Kenya, Nepal, Mozambique, Bangladesh, Cambodia, Lebanon, Tajikistan, and the Philippines (2017--2022). Target: observed number of ARI cases among children under five per cluster. Baseline: gaseous pollutants (CO, NO\textsubscript{2}, SO\textsubscript{2}) plus controls (elevation, population, temperature, urban/rural). Feature augmentation: 64-dim AlphaEarth embedding sampled at each cluster's survey year over a 2--5 km buffer. Evaluation: 5-fold $\times$ 2-repeat cross-validation, tested with three tree-based estimators.

\begin{figure}[H]
\centering
\begin{subfigure}[t]{0.70\textwidth}
    \centering
    \includegraphics[width=\textwidth]{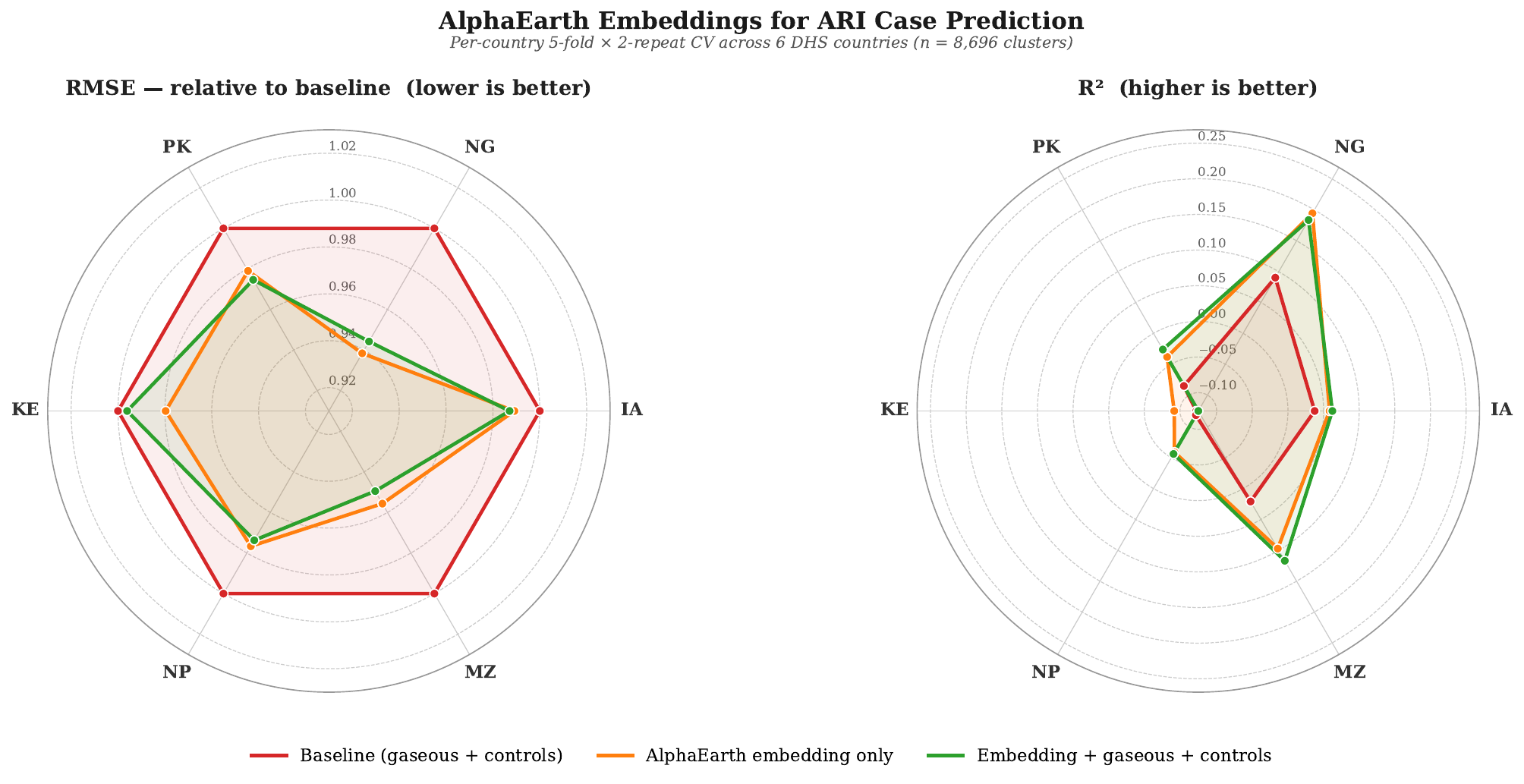}
    \caption{\textbf{Per-country performance for the six DHS countries with $n \geq 139$ clusters} (India, Pakistan, Nigeria, Kenya, Nepal, Mozambique; 8{,}696 of 9{,}271 clusters total). Left panel: RMSE relative to baseline (lower is better). Right panel: test $R^2$ (higher is better). The three polygons compare the gas+controls baseline (red), AlphaEarth alone (orange), and the combined model (green). The combined model strictly dominates baseline on every spoke; the largest absolute gains are in Mozambique ($\Delta R^2 = +0.096$) and Nigeria ($+0.093$).}
    \label{fig:case2_radar}
\end{subfigure}
\hfill
\begin{subfigure}[t]{0.50\textwidth}
    \centering
    \includegraphics[width=\textwidth]{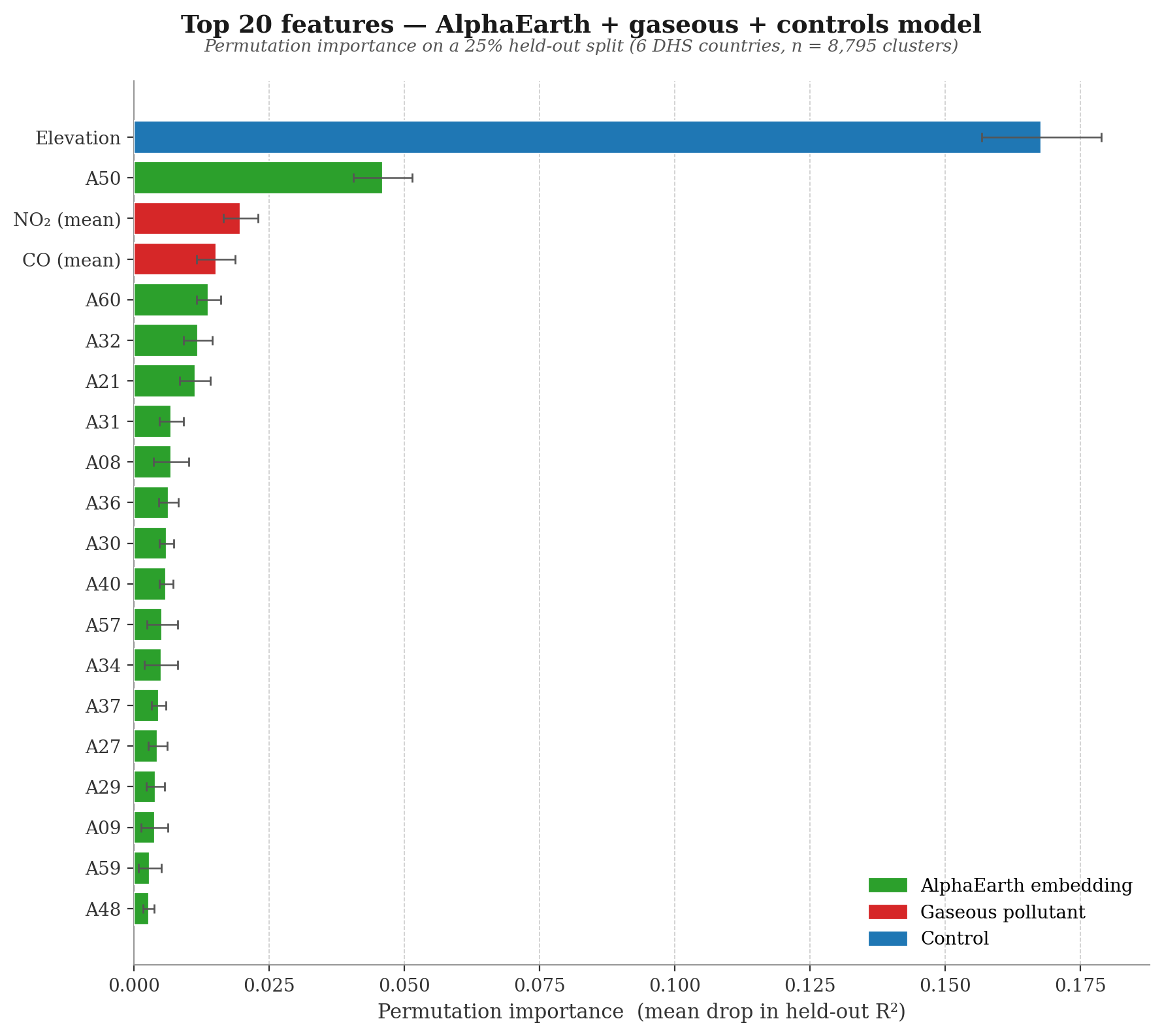}
    \caption{\textbf{Top-20 permutation importances on a 25\% held-out split} (combined embedding + gaseous pollutants + controls model). Bars coloured by feature group: AlphaEarth (green), gaseous pollutants (red), controls (blue). Elevation leads, but seven of the top ten features are AlphaEarth bands, and two of the AlphaEarth bands (A50, A60) outrank the strongest pollutant (NO\textsubscript{2}).}
    \label{fig:case2_bar}
\end{subfigure}
\caption{\textbf{Case 2 --- Childhood ARI prediction across 11 LMICs (DHS, 2017--2022; 5-fold $\times$ 2-repeat CV).} The pooled $R^2$ improves from 0.157 (gaseous pollutants + controls) to 0.206 (gas + controls + AlphaEarth). The improvement is consistent across three tree-based estimators (Random Forest, HistGradientBoosting, XGBoost), indicating that the signal lies in the embeddings rather than in any one model's inductive bias.}
\label{fig:case2}
\end{figure}

\paragraph{Result.} At the pooled level (Figure~\ref{fig:case2_ablation-global} in appendix), all three estimators show a consistent monotonic improvement: baseline $R^2 \approx 0.15$--0.17 $\rightarrow$ AlphaEarth-only $\approx 0.18$--0.19 $\rightarrow$ both features combined $\approx 0.21$. The cross-model consistency of the effect (best: XGBoost, $R^2=0.210$) indicates the signal is in the embeddings, not an artefact of any particular inductive bias. Per-country (Figure~\ref{fig:case2_radar}), embeddings help in all six countries with $n \geq 139$ clusters (8{,}696 of 9{,}271 clusters, 94\% of the data), with largest absolute gains in Mozambique ($\Delta R^2 = +0.096$) and Nigeria ($+0.093$). In the smallest-$n$ countries ($n < 130$) the 64-dim embedding causes overfitting under HistGradientBoosting; Random Forest substantially attenuates this and is recommended in data-scarce settings.

\section{Case 3 --- Stunting (WHZ), 35 DHS Countries}

\paragraph{Setup.} 389{,}035 children under five across 35 DHS countries surveyed 2015+, with tabular covariates (temperature, age, altitude, household income, Composite Coverage Indicator, sanitation, sex) predicting country-demeaned Weight-for-Height Z-score. Model: 3-layer MLP with an 8-dim country embedding, trained with Huber loss and DHS sample weights. Baseline achieves test $R^2 = 0.036$; fixed-effects regression with the same specification achieves $R^2 = 0.014$ on the identical cluster-held-out fold.

\begin{figure}[H]
\centering
\begin{subfigure}[t]{0.50\textwidth}
    \centering
    \includegraphics[width=\textwidth]{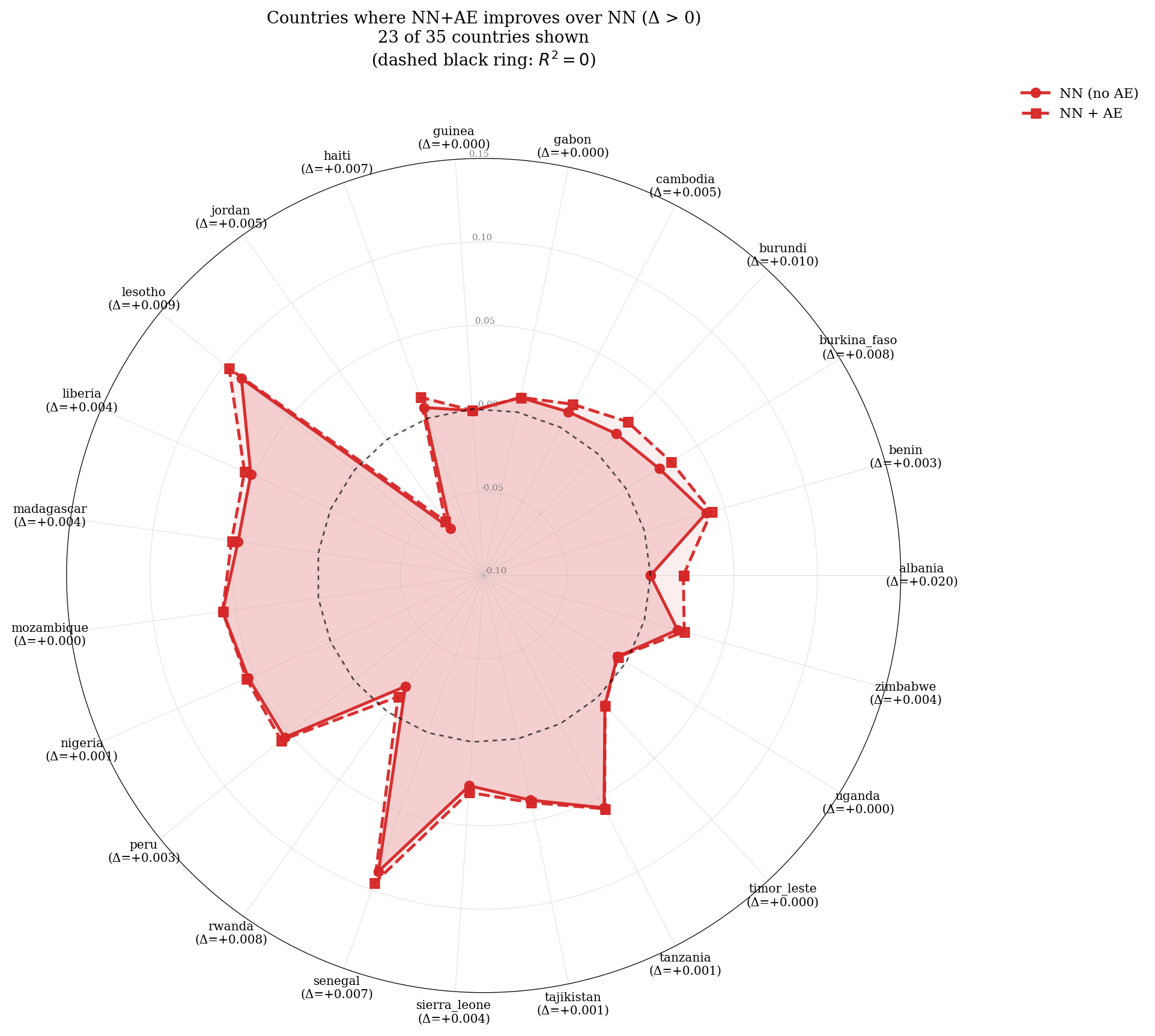}
    \caption{\textbf{Per-country test $R^2$ for the 23 of 35 countries where the country-broadcast AE input yielded $\Delta R^2 > 0$.} Solid polygon: NN baseline. Dashed polygon: NN + country-broadcast AE. The two polygons are visually indistinguishable (max gain $\Delta R^2 = +0.020$ in Lesotho; no country exceeds the $\Delta = +0.01$ substantive threshold), confirming that AE provides no useful information once a country embedding is already in the model.}
    \label{fig:case3_radar}
\end{subfigure}
\hfill
\begin{subfigure}[t]{0.80\textwidth}
    \centering
    \includegraphics[width=\textwidth]{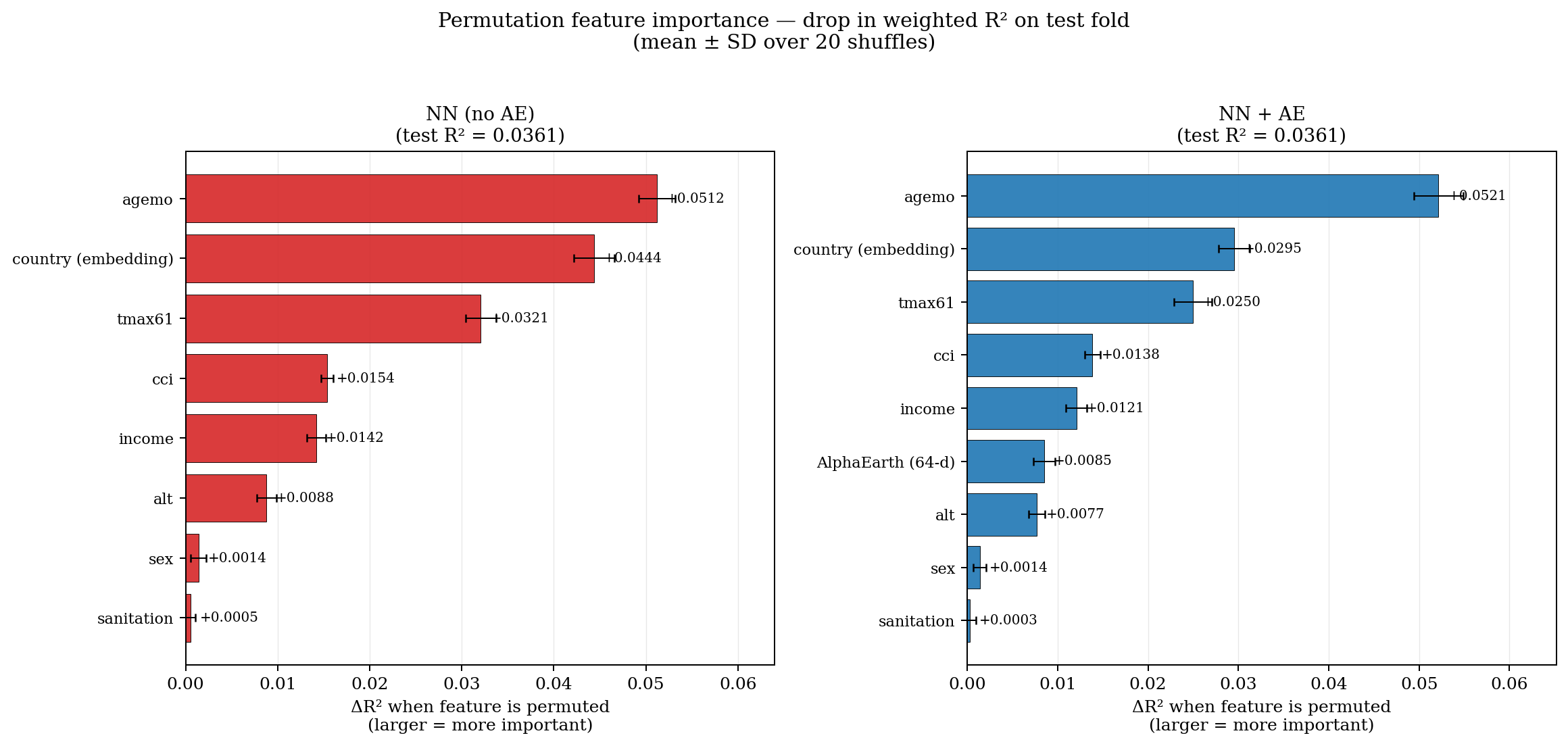}
    \caption{\textbf{Permutation feature importance (drop in test-fold weighted $R^2$, mean $\pm$ SD over 20 shuffles).} Left: NN baseline. Right: NN + country-broadcast AE. When AE is added, importance partially shifts from the country embedding (0.044 $\rightarrow$ 0.030) onto the AE input (0.009), confirming the collinearity --- AE absorbs information already carried by country identity rather than adding new signal.}
    \label{fig:case3_bar}
\end{subfigure}
\caption{\textbf{Case 3 --- Stunting (WHZ) prediction across 35 DHS countries, 2015+ (country-level AE prototype).} A country-broadcast AlphaEarth fingerprint is constant within each country and therefore collinear with the model's country fixed effect / embedding; as predicted, it yields no meaningful gain ($\Delta R^2 \approx 0$). This negative result motivates the cluster-level extraction (described in the text) as the experiment that can actually test AE's contribution to undernutrition prediction.}
\label{fig:case3}
\end{figure}

\paragraph{Prototype test (country-level AlphaEarth).} We extracted 30 population-weighted points per country, averaged to a country-level fingerprint, and broadcast this 64-dim vector to every child in that country as an additional NN input. Figures~\ref{fig:case3_radar} and~\ref{fig:case3_bar} show the result: country-broadcast AE adds at most $\Delta R^2 = +0.020$ for any country, and no country clears a substantive $\Delta = +0.01$ threshold. This is the expected result: a feature that is constant within each country cannot add information beyond the country fixed effect or learned country embedding that is already in the model. We show it here because it confirms, through direct measurement, the structural argument that \emph{cluster-level} AE extraction is necessary for this outcome.

\paragraph{The target experiment.} The analytical sample contains 45{,}775 unique DHS clusters. A cluster-level AlphaEarth extraction would provide $\sim\!1{,}000\times$ more fingerprints than the country-level prototype, with within-country variation (urban vs rural, coastal vs inland, forest vs dryland) that is structurally absent at country scale. Based on Cases 1 and 2, we expect cluster-level AE to carry real predictive signal for stunting; we are currently awaiting DHS cluster coordinate access from collaborators to run this experiment.

\section{Synthesis: Where AlphaEarth Helps}

Across these three outcomes a consistent pattern emerges, summarised in Table~\ref{tab:synth}.

\begin{table}[h!]
\centering
\small
\begin{tabular}{lllll}
\toprule
Outcome & Grain & Baseline $R^2$ & + AE $R^2$ & $\Delta R^2$ \\
\midrule
Malaria (Nigeria)              & Cluster (28{,}209)        & 0.201 & 0.245 & $+0.044$ \\

ARI (11 LMICs, pooled)         & Cluster (9{,}271)          & 0.157 & 0.206        & $+0.049$ \\
WHZ (35 LMICs)                 & Country (broadcast)        & 0.036 & 0.036        & $\approx 0$ \emph{(expected)} \\
WHZ (planned)                  & Cluster ($\sim$45{,}775)   & 0.036 & ?            & (blocked on data access) \\
\bottomrule
\end{tabular}
\caption{AlphaEarth contributions across the three studies. The spatial grain of the merge determines whether the embedding can contribute --- country-level broadcast is collinear with country identity, cluster-level is informative.}
\label{tab:synth}
\end{table}

The embeddings add value when they can express \emph{within-country} variation --- at the cluster or location level. When forced to the country level (Case 3 prototype), they are absorbed by country fixed effects and cannot contribute. Cases 1 and 2 both operate at cluster/location granularity and both show the expected gains; Case 3 is blocked on cluster-level geodata that is currently not in hand.

\section{Conclusion}
Across three structurally distinct health outcomes in LMICs, AlphaEarth Foundations satellite embeddings consistently add predictive value when applied at cluster or location level. The gains in malaria prediction ($\Delta$ $R^2$=+0.044) and childhood ARI prediction ($\Delta$ $R^2$=+0.049) are geographically uniform and robust across model classes, supporting the interpretation that static landscape structure encodes transmission and exposure signals not captured by conventional climate or pollution covariates. The country-level prototype for stunting behaves as the structural argument predicts: collinearity with country fixed effects renders it uninformative, and the cluster-level experiment remains the definitive test. These results establish AlphaEarth embeddings as a promising and generalisable feature class for population health modelling in low-resource settings, with clear implications for disease surveillance, early warning systems, and the targeting of public health interventions.

\section{Limitations}
Several limitations should be noted. First, DHS cluster coordinates are displaced by up to 5 km (urban) or 10 km (rural) for privacy protection~\citep{burgert2013}; this introduces measurement error in the spatial merge that is likely to attenuate, rather than inflate, the observed embedding effects. Second, AlphaEarth embeddings are static annual composites and cannot capture intra-annual environmental dynamics; monthly or seasonal embeddings would be needed to model outbreak-scale variation. Third, the stunting analysis relies on Weight-for-Height Z-score as a proxy for child undernutrition; stunting (height-for-age) and wasting (weight-for-height) are related but distinct, and the pipeline's performance on height-for-age Z-score has not yet been evaluated. Finally, the country-level pooling in the ARI and WHZ analyses may mask heterogeneity in embedding utility across different agroecological zones and health system contexts; within-country analyses at finer spatial scales remain an important direction for future work.

\section{Future work}

We see three concrete ways the Google Earth AI model could accelerate this line of work:

\begin{enumerate}
\item \textbf{Population Dynamics Foundation embeddings.} The Population Dynamics model includes health and socio-demographic indicators that would complement the landscape-focused AlphaEarth representation. A combined feature set would be particularly valuable for stunting (where nutritional status depends on both environmental and structural determinants) and for ARI (where health-system covariates partially compete with pollution features in the baseline). Direct access to pre-extracted Population Dynamics embeddings matched to DHS cluster coordinates would enable a three-way ablation (tabular only / + AlphaEarth / + AlphaEarth + Population Dynamics) analogous to what Case 1 demonstrates for single-modality embeddings.

\item \textbf{Cluster-level AlphaEarth at scale.} Our Case 3 extraction of 30 points $\times$ 35 countries is a prototype; the full experiment requires $\sim\!45{,}775$ cluster-centred extractions with 2--5 km buffers matching DHS privacy displacement. A batched extraction pipeline or direct access to pre-computed cluster-level fingerprints for the DHS program would substantially shorten the time to a definitive result.

\item \textbf{Methodological feedback.} We would value the AE team's perspective on two specific choices: (a) whether to weight within-buffer pixels uniformly or by population density during cluster-level averaging, and (b) whether multi-year embedding composites (e.g. 3-year means) or survey-year point estimates are preferable for linking to DHS surveys with staggered fieldwork.
\end{enumerate}


\bibliographystyle{unsrtnat}
\bibliography{references}

\begin{appendix}
\section{Case 1 --- Malaria Prediction, Nigeria}\label{case1_resnet}

\begin{figure}[H]
\centering
\begin{subfigure}[t]{0.40\textwidth}
    \centering
    \includegraphics[width=\textwidth]{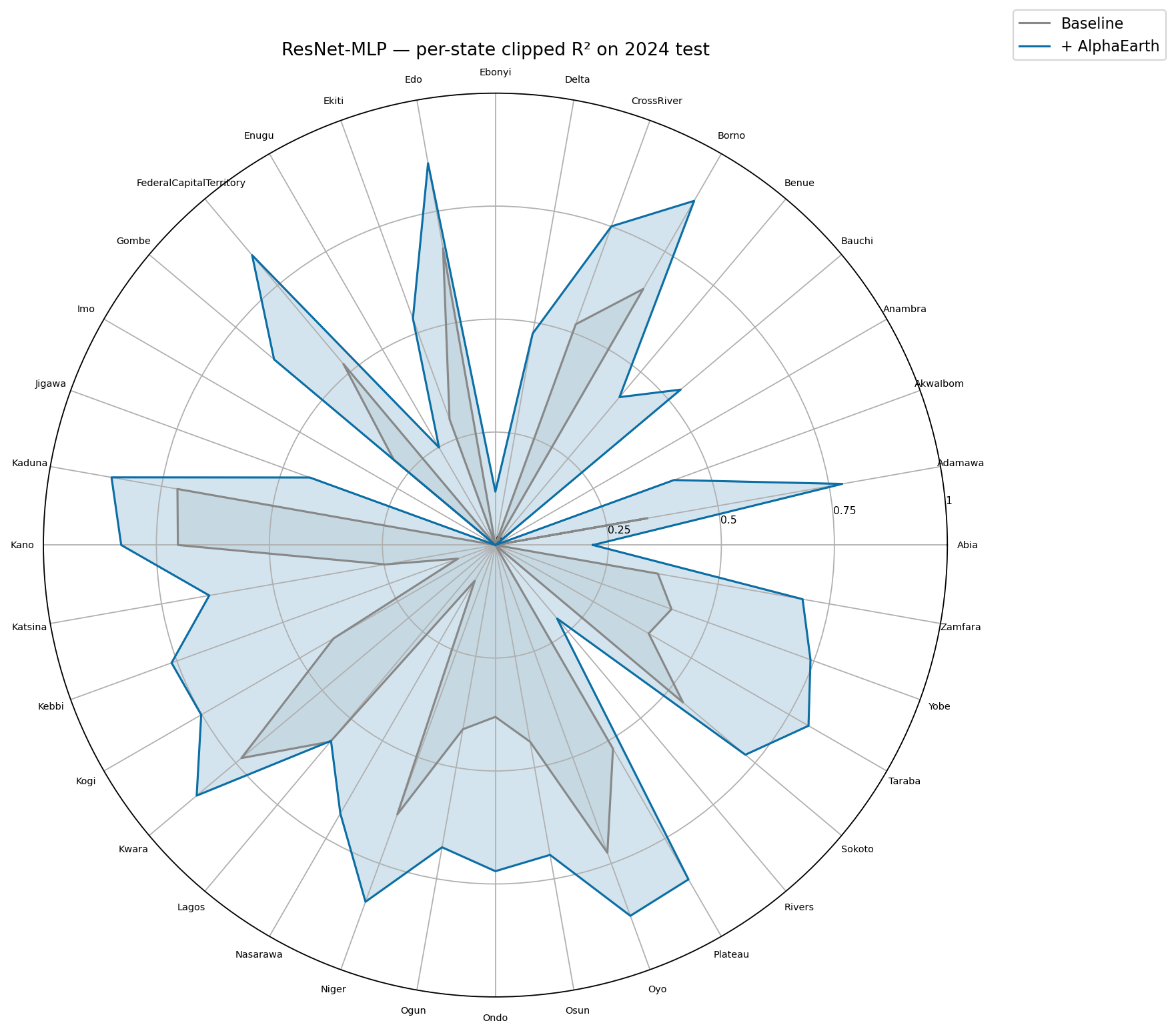}
    \caption{\textbf{Per-region 2024 test $R^2$ across Nigerian states.} Each spoke is one state; the inner polygon is the climate-only baseline and the outer polygon is the same model with the 64-dim AlphaEarth fingerprint appended. The outer polygon strictly dominates on every spoke, indicating that the lift is geographically uniform rather than driven by a few high-burden states.}
    \label{fig:case1_radar_resnet}
\end{subfigure}
\hfill
\begin{subfigure}[t]{0.40\textwidth}
    \centering
    \includegraphics[width=\textwidth]{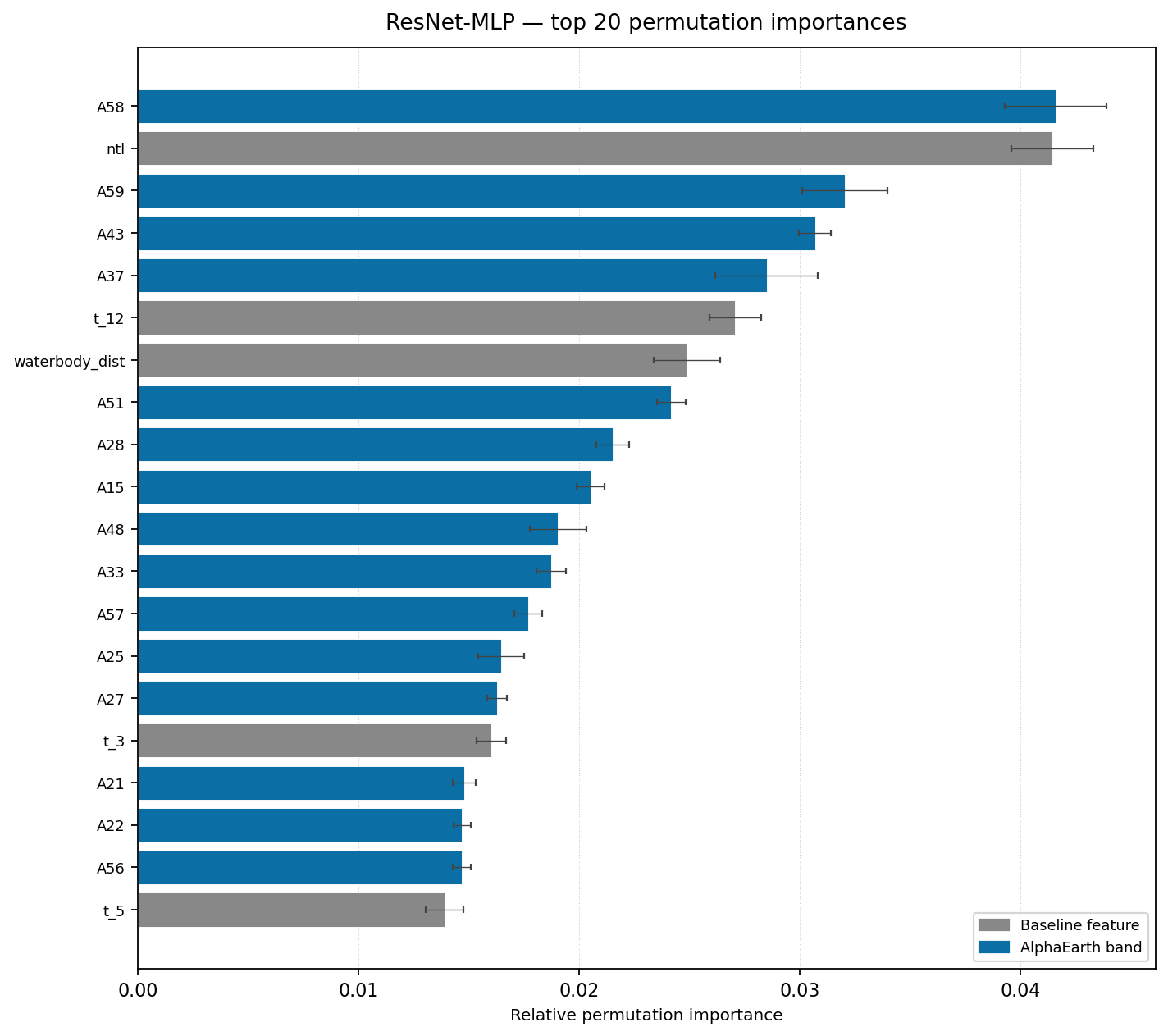}
    \caption{\textbf{Top-20 ResNet feature importances.} Bars are coloured by feature group (baseline climate / contextual vs. AlphaEarth band $A_i$).}
    \label{fig:case1_bar}
\end{subfigure}
\caption{\textbf{Case 1 --- Malaria case prediction in Nigeria (NMEP, 2000--2024; train 2000--2023, test 2024).} AlphaEarth embeddings provide a geographically uniform $R^2$ gain (left) and emerge as the dominant feature group in the importance decomposition (right), supporting the interpretation that static landscape structure carries malaria-transmission signal not captured by monthly climate covariates.}
\label{fig:case1}
\end{figure}

\newpage

\section{Case 2 --- Childhood ARI, 11 DHS Countries}
\begin{figure}[H]
\centering
\includegraphics[width=0.70\linewidth]{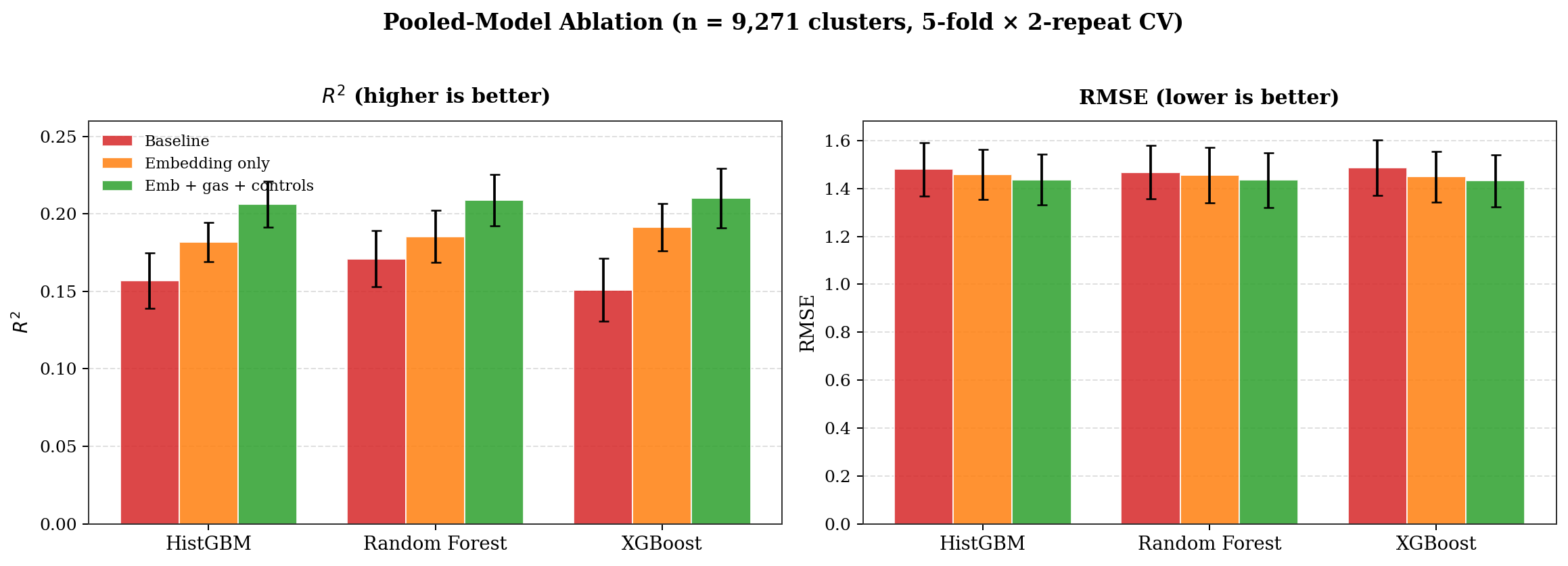}
\caption{Pooled-model ablation. $R^2$ (left) and RMSE (right) for
each estimator--feature-set combination. Error bars denote
cross-fold standard deviation across 10 folds. All three models
agree that adding embeddings to the gaseous baseline yields a
consistent improvement, with XGBoost reaching the highest
$R^2 = 0.210$ and Random Forest the lowest RMSE.}
\label{fig:case2_ablation-global}
\end{figure}

\end{appendix}

\end{document}